\documentclass[a4paper,12pt]{article}
\pdfoutput=1 

\usepackage{jcappub} 
\usepackage{subfigure}
\usepackage[T1]{fontenc} 
\usepackage{lipsum}
\usepackage{amsmath, multirow, amssymb, wasysym, gensymb}
\usepackage{cleveref}

\title{\boldmath The Dark Sequential $Z^\prime$ Portal: Collider and Direct Detection Experiments}

\author[a]{Giorgio Arcadi,}
\author[a]{Miguel D. Campos,}
\author[a]{Manfred Lindner,}
\author[b,c]{Antonio Masiero,}
\author[a]{Farinaldo S. Queiroz}

\affiliation[a]{Max-Planck-Institut f\"ur Kernphysik, Saupfercheckweg 1, 69117 Heidelberg,
Germany}
\affiliation[b]{Dipartimento di Fisica e Astronomia ``G. Galilei'', Universita di Padova, Italy}
\affiliation[c]{Istituto Nazionale Fisica Nucleare, Sezione di Padova, I-35131 Padova, Italy}

\emailAdd{miguel.campos@mpi-hd.mpg.de}
\emailAdd{arcadi@mpi-hd.mpg.de}
\emailAdd{masiero@pd.infn.it}
\emailAdd{queiroz@mpi-hd.mpg.de}

\abstract{ 
We revisit the status of a Majorana fermion as a dark matter candidate when a sequential $Z^\prime$ gauge boson dictates the dark matter phenomenology. Direct dark matter detection signatures rise from dark matter-nucleus scatterings at bubble chamber and liquid xenon detectors, and from the flux of neutrinos from the Sun measured by the IceCube experiment, which is governed by the spin-dependent dark matter-nucleus scattering. On the collider side, LHC searches for {\it dilepton} and {\it mono-jet + missing energy} signals play an important role. The relic density and perturbativity requirements are also addressed. By exploiting the dark matter complementarity we outline the region of parameter space where one can successfully have a Majorana dark matter particle in light of current and planned experimental sensitivities.
}

\begin{document}
\maketitle
\flushbottom

\section{Introduction}
\label{sec:intro}

The existence of Dark Matter (DM) is a fact that has been accumulating evidence since the early 70's (or even before if we consider the analysis done by Franz Zwicky in 1933 that led to him coining the term \textit{Dunkle Materie} \citep{Zwicky:1933gu}). However its true nature remains an open question in physics as of today. Among the particle candidates Weakly Interacting Massive Particles (WIMPs) stand out, for being able to reproduce the observed relic abundance in a rather natural way and predict signals at current or planned experiments \cite{Olive:2003iq}. Despite this theoretical motivation and intense experimental efforts, the existence of WIMPs has not yet been established (for a recent review see \citep{Arcadi:2017kky}), for which distinct candidates and their possible signals in different detectors must be explored. \\

\noindent
The experimental efforts can be roughly classified in three: indirect detection \cite{Jungman:1995df,Bertone:2004pz}, direct detection \cite{Feng:2010gw,Klasen:2015uma,Undagoitia:2015gya,Queiroz:2016sxf,Kavanagh:2017hcl} and collider searches \cite{Abdallah:2015ter,Kahlhoefer:2017dnp,Plehn:2017fdg} having each of these generally complementary characteristics. Although, depending on the details of the model, the different search strategies are not equally effective. Indeed, in the scenario investigated here, Majorana fermion dark matter, indirect searches performed by experiments like Fermi-LAT, MAGIC and H.E.S.S. cannot probe the parameter space corresponding to the viable DM relic density \cite{Aharonian:2006wh,Barger:2009yt,Abazajian:2011ak,Cembranos:2012nj,Ackermann:2015lka,Abdallah:2016ygi,Charles:2016pgz,Ahnen:2016qkx,Beck:2017hkp,Fermi-LAT:2016uux,TheFermi-LAT:2017vmf}. Even with the Cherenkov Telescope Array, indirect detection probes are bound to be sub-dominant \cite{Bringmann:2008kj,Pierre:2014tra,Silverwood:2014yza,Ibarra:2015tya,Roszkowski:2016bhs,Balazs:2017hxh}.~\footnote{It has been noted that Majorana particles mediated by a charged scalar can yield observable indirect detection signatures via internal bremsstrahlung, but this is not the scenario under study \cite{Bringmann:2012vr,Bringmann:2012ez}.} This is because the s-wave (i.e. velocity independent) component of its annihilation cross-section into SM fermions is helicity suppressed. On the other hand, even if not capable of significantly  influence the flux of gamma-rays/cosmic rays, DM annihilation processes occurring at present times can effectively influence the flux of neutrinos from the Sun, detectable by neutrino telescopes such as IceCube. This is due to the large exposure of these detectors to the Sun and, more importantly, because the neutrino flux is dictated by the unsuppressed WIMP-nucleon scattering cross-section. For these reasons neutrinos from the Sun offer a complementary probe with respect to Earth based experiments~\cite{Bergstrom:1996kp,Gondolo:2004sc,Blennow:2007tw,Wikstrom:2009kw,Lattanzi:2014mia,Yaguna:2016bga}.\\

\noindent
In particular, we will perform a detailed study of the phenomenology of a Majorana DM candidate interacting with a spin-1 mediator dubbed $Z'$. For simplicity, and to minimize the number of free parameters, we will assume that the $Z'$ couples with the SM fermions in the exact same way as the SM $Z$-boson~\cite{Alves:2013tqa}.~\footnote{For other constructions in the context of Majorana dark matter see \cite{Kouvaris:2007iq,Belotsky:2008vh,Goodman:2010yf,Garny:2015wea,Ciafaloni:2011sa,Ciafaloni:2011gv,Duerr:2015vna,Heisig:2015ira,Dutra:2015vca,Anchordoqui:2015fra,Chua:2015ixi,Hooper:2014fda,Nomura:2016vxr,Chao:2016avy,Pires:2016vek,Matsumoto:2016hbs}.} This setup is also referred as Sequential Standard Model (SSM). Alternative assignations of these couplings can be motivated by identifying the $Z'$ with the gauge boson of an additional, with respect to the SM gauge group, $U(1)$ symmetry~\cite{Alvares:2012qv,Profumo:2013sca,Alves:2015pea,Alves:2015mua,Duerr:2015wfa,Allanach:2015gkd,Lindner:2016lpp,Klasen:2016qux,Altmannshofer:2016jzy,Alves:2016fqe,Alves:2016cqf,Arcadi:2017atc,Campos:2017dgc}. Given our assumptions, the model has only three free parameters, being the DM and $Z'$ masses and the coupling of the DM with the $Z'$.\\

\noindent
The important observables in this minimized setup are the dark matter relic density, the spin-dependent WIMP-nucleon scattering cross section, and the $Z^\prime$ production rate at the LHC. The dark matter relic density is computed in the usual thermal equilibrium framework leading to a freeze-out governed by the dark matter annihilation cross section into SM fermions. The direct dark matter detection signatures stem from spin-dependent dark matter-nucleus scatterings at Bubble Chamber and Liquid XENON detectors and from dark matter annihilations in the Sun. As for colliders, LHC searches for signal events in dilepton and mono-jet channel provide restrictive bounds on the model. Both probes are highly sensitive to the $Z^\prime$ production cross section at the LHC. \\

\noindent
That said, we exploit the complementarity between these observables to outline the viable region of parameter where one can successfully have a Majorana dark matter particle in the context of the sequential dark $Z^\prime$ portal.~\footnote{We emphasize that we are concerned just with the DM phenomenology, and we will not explore the possibilities or limitations of the additional $U(1)$ symmetry and consider just the constraints coming from DM searches. }\\

\noindent
Our work is organized as follows: in \cref{sec:Zp} we describe the Majorana dark matter model we investigate; in \cref{sec:constraints} we introduce the observables and experimental constraints; in \cref{sec:res} we summarize and discuss our finding. Finally, in \cref{sec:conclusions} we draw our conclusions.\\

\section{The Dark Sequential $Z^\prime$ portal}
\label{sec:Zp}

Seen usually as natural consequences of a symmetry breaking chain in Grand Unified Theories (GUTs) \citep{Arcadi:2017atc} and many other extended gauge sectors \cite{Mizukoshi:2010ky,Alvares:2012qv,Profumo:2013sca,Kelso:2013nwa,Heeck:2014qea,Rodejohann:2015lca,Patra:2015bga,Patra:2016ofq,Lindner:2016lxq,Lindner:2016lpp}, $U(1)$ groups are ubiquitous in high energy physics model building for being the simplest continuous Abelian group available. The breaking down to the SM group typically leads to a massive gauge boson. If the SM Higgs doublet is not charged under the new $U(1)$ group and the $U(1)$ symmetry is spontaneously broken via a scalar singlet then there will be no mass mixing between the $Z$ and $Z^\prime$ gauge bosons. In this kind of framework the $Z'$ represents the only ``portal'' between the DM and the SM fermions.~\footnote{We will consider the case in which the $Z'$ has direct coupling with the SM fermions. Even if this would not be the case if a coupling between the $Z'$ and the SM is originated by the Lorentz and gauge invariant kinetic mixing term $\delta B^{\mu \nu}B_{\mu \nu}^{'}$. We will not consider here this kind of scenario. As pointed in the last section it is nevertheless possible to straightforwardly generalize our results.} Spin-1 portals are also among the most adopted benchmarks for collider searches of Dark Matter. In this case, however, ``simplified'' models in which the $Z'$ interacts only with quarks are customarily considered, see however~\cite{Albert:2017onk}. These kinds of setups are contrived from the theoretical point of view~\cite{Kahlhoefer:2015bea,Ellis:2017tkh} and do not account for the relevant impact of collider searches for dilepton resonances. \\

\noindent
In order to have a simple and predictable setup, which can be at the same time easily extended towards more motivated frameworks, we will consider the case of a Majorana fermion coupled with a sequential $Z'$ boson, and then refer to it as ``Sequential Dark $Z^\prime$ Portal''. 
The relevant part of the Lagrangian then looks like,

\begin{equation}\label{eq:L}
\mathcal{L}\supset\left[ g_\chi {\chi}\gamma^\mu\gamma^5 \chi + \sum_{f\in \text{SM}}\bar{f}\gamma^\mu(g_{fv}+g_{fa}\gamma^5)f \right]Z^\prime_\mu,
\end{equation} where the sum is over all the SM fermions and the factors $g_{fv}$ and $g_{fa}$ are given by,

\begin{equation} \label{eq:g}
\begin{aligned}
g_{uv}=\dfrac{-e}{4}\left( \dfrac{5}{3} \tan\theta_w - \cot\theta_w \right),\quad & g_{ua}=\dfrac{-e}{4}\left( \tan\theta_w + \cot\theta_w \right)\\
g_{dv}=\dfrac{e}{4}\left( \dfrac{1}{3} \tan\theta_w - \cot\theta_w \right),\quad & g_{da}=\dfrac{e}{4}\left( \tan\theta_w + \cot\theta_w \right)\\
g_{\ell v}=\dfrac{e}{4}\left(3 \tan\theta_w - \cot\theta_w \right),\quad & g_{\ell a}=\dfrac{e}{4}\left( \tan\theta_w + \cot\theta_w \right)\\
g_{\nu v}=\dfrac{e}{4}\left(\tan\theta_w + \cot\theta_w \right),\quad & g_{\nu a}=\dfrac{-e}{4}\left( \tan\theta_w + \cot\theta_w \right)
\end{aligned}
\end{equation}

\noindent
with $u$, $d$, $\ell$ and $\nu$ the up-type, down-type quarks, charged leptons and neutrinos respectively, $e=\sqrt{4\pi \alpha}$ is the electromagnetic coupling and $\theta_w$ the Weinberg angle.\\

\noindent
The Lagrangian in \cref{eq:L} generates two kind of operators relevant for direct detection of DM: $\bar{f}\gamma^\mu f\chi\gamma_\mu\gamma_5\chi$ and $\bar{f}\gamma^\mu\gamma_5 f\chi\gamma_\mu\gamma_5\chi$. The former yields a spin independent (SI), whereas the latter a spin dependent (SD) interaction, respectively. Since the SI cross section is velocity suppressed (being the scattering cross section proportional to $v^2 \sim 10^{-6}$) we will concentrate on the constraints coming exclusively from SD searches. \\

\noindent
Now that we have set up the framework to investigate, we will discuss the relevant observable and respective constraints applicable to the model.\\

\section{Experimental constraints}
\label{sec:constraints}

\subsection{Relic Density}
\begin{figure}[h]
\centering
\subfigure{
\includegraphics[height=0.1\textheight]{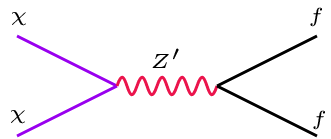} 
}
\subfigure{
\includegraphics[height=0.1\textheight]{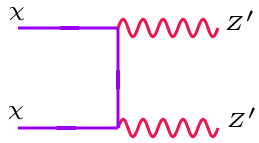}
}
\caption{\label{fig:diagram} 
Feynman diagrams relevant for dark matter annihilation. The first encompasses all possible annihilations into SM particles through the $Z^\prime$ portal, whereas the second the self-annihilation into $Z^\prime$ gauge bosons.
}
\end{figure}

The first requirement that we will impose upon the framework presented is that the Majorana fermion reproduces the observed relic abundance of DM, namely $\Omega h^2\approx 0.12$ \citep{Ade:2015xua} through thermal production. The annihilations channels relevant for this production are presented in \cref{fig:diagram} and correspond to an s-channel annihilation mediated by a $Z^\prime$ and a t-channel process $\chi\chi\to Z^\prime Z^\prime$ when kinematically allowed.\\

\noindent
The DM relic density has been precisely determined in numerical way through the package \texttt{MicrOMEGAs 4.3.2}~\cite{Barducci:2016pcb}. Useful analytical approximations are nevertheless provided by the velocity expansion (see also~\citep{Alves:2016cqf}):

\begin{align}
& \langle \sigma v \rangle_{ff}=\sum_{f} n_c^f
 \frac{2\sqrt{m_\chi^2-m_f^2}}{\pi m_\chi M_{Z'}^4 
\left(M_{Z'}^2-4 m_\chi^2\right)^2} 
\left[{(g_{fa})}^2 g_\chi^2 m_f^2 \left(M_{Z'}^2-4 m_\chi^2\right)^2 \right]\nonumber\\
&-\frac{v^2}{6 \pi  m_\chi M_{Z'}^4 \sqrt{m_\chi^2-m_f^2} \left(M_{Z'}^2-4 m_\chi^2\right)^3} 
\left[{(g_{fa})}^2 \left\{-g_\chi^2 
\left(M_{Z'}^2-4 m_\chi^2\right)\right.\right.\nonumber\\
& \left.\left. \times \left(23 m_f^4 M_{Z'}^4-192 m_f^2 m_\chi^6-4 m_f^2 m_\chi^2 M_{Z'}^2 \left(30 m_f^2+7
   M_{Z'}^2\right)\right.\right.\right.\nonumber\\
& \left.\left.\left. +8 m_\chi^4 \left(30 m_f^4+12 m_f^2 M_{Z'}^2+M_{Z'}^4\right)
\right)\right\}\right.\nonumber\\
&\left.+M_{Z'}^4 {(g_{fv})}^2 \left\{4 g_\chi^2 
\left(m_f^4+m_f^2 m_\chi^2-2 m_\chi^4\right) \left(M_{Z'}^2-4 m_\chi^2\right)\right\}\right];
\label{Eq:sigvff2}
\end{align}	

\begin{align}
& \langle \sigma v \rangle_{Z'Z'}=\frac{g_\chi^4}{\pi m_\chi^2}{\left(1-\frac{M_{Z'}^2}{m_\chi^2}\right)}^{\frac{3}{2}}{\left(1-\frac{M_{Z'}^2}{2 m_\chi^2}\right)}^{-2}\nonumber\\
&+ \frac{g_\chi^4 v^2}{3 \pi m_\chi^2}\sqrt{1-\frac{M_{Z'}^2}{m_\chi^2}}{\left(1-\frac{M_{Z'}^2}{2 m_\chi^2}\right)}^{-4}\left(\frac{23}{16}\frac{M_{Z'}^6}{m_\chi^6}-\frac{59}{8}\frac{M_{Z'}^4}{m_\chi^4}+\frac{43}{4}\frac{M_{Z'}^2}{m_\chi^2}+2-12 \frac{m_{\chi}^2}{M_{Z'}^2}+8 \frac{m_{\chi}^4}{M_{Z'}^4}\right).
\label{Eq:sigvZpZp}
\end{align}

\noindent
where $n_c^f$ is the color factor and the $g_{fv},g_{fa},f=u,d,e,\mu,\tau,\nu$ have been defined in \cref{eq:g}. In the first expression the sum runs over the final states kinematically accessible for a given value of the DM mass $m_\chi$.\\

\noindent
Some important features are worth noticing in the expressions above. Concerning the annihilation into fermions we see that the s-wave (velocity independent term) is proportional to $\frac{m_f^2}{m_{Z'}^4}$ with $m_f$ being the final state fermion mass. Hence, unless annihilation into top quarks is kinematically accessible, the s-wave term of the DM annihilation cross-section is strongly suppressed so that the dominant contribution comes from the p-wave term that does exhibit the $Z^\prime$ resonance. Because of this, there is a strong mismatch between the value of the annihilation cross-section at thermal freeze-out, relevant for the relic density, corresponding to $v \sim 0.3$, and the one at present times, possibly relevant for an indirect detection signal, corresponding instead to $v \sim 10^{-3}$. \\

\noindent
The reasoning above explains why indirect dark matter detections is bound to be sub-dominant in this scenario, because the s-wave annihilation cross section which is relevant for indirect dark matter searches is helicity suppressed, only becomes sizable when $m_{\chi} > m_{top}$, i.e. when $m_{\chi} \simeq 200$~GeV, but indirect detection limits for $m_{\chi} > 200$~GeV are not strong. We remind the reader that would not occur in the case of a Dirac fermion since the s-wave component of the annihilation cross-section is not be helicity suppressed~\cite{Mizukoshi:2010ky,Profumo:2013sca,Alves:2013tqa}.\\

\noindent
When the dark matter mass becomes larger than the $Z^\prime$ mass the annihilation to $Z^\prime$ pairs opens up, and \cref{Eq:sigvZpZp} becomes relevant. Notice that \cref{Eq:sigvZpZp} features s-wave and p-wave contributions. The s-wave term scales with $1/m_\chi^2$, so one can naively expect that this contribution is small since we are dealing with a heavy DM particle, with $m_{\chi} > m_{Z^\prime}$ term. However, the p-wave term grows with $m_\chi^2/M_{Z^{'}}^4$. Since we are now focused on a heavy DM particle, this enhancement with $m_\chi^2$ compensates the $v^2$ suppression and dominates the overall DM annihilation. Such enhancement with $m_\chi^2/M_{Z^{'}}^4$ is related to the annihilation into longitudinal $Z^{'}$ and is actually pathological, due to the non-UV complete framework we investigate. We will discuss this point in further detail in \cref{ssec:perturbativity}.

\begin{figure}[h]
\centering
\subfigure{
\includegraphics[height=0.1\textheight]{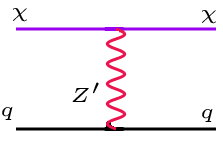} 
}
\caption{\label{fig:diagramDD} 
Feynman diagram relevant for direct detection. The dark matter scattering off nucleons occurs via the $Z^\prime$ t-channel exchange.
}
\end{figure}

\subsection{Direct Detection}
\label{ssec:DD}

Direct detection signatures of the model rely on the SD dark matter-scattering off nucleons via the t-channel $Z^\prime$ exchange as displayed in \cref{fig:diagramDD}. It is dictated by the axial couplings $g_{fa}$ defined in \cref{eq:g}, and described by the following cross-section,
\begin{equation}
\label{eq:SD}
\sigma_{\chi N}^{\rm SD}= \frac{12 \mu_{\chi p}^2}{\pi}\frac{g_\chi^2}{M_{Z'}^4}{\left[ g_{ua}\Delta_u^N+g_{da}\left(\Delta_d^N+\Delta_s^N\right) \right]}^2,
\,\,N=p,n,
\end{equation}while, on the contrary, sizeable Spin Independent (SI) interactions are absent since Majorana fermions have null vectorial couplings. We now address the bounds considered in this work.\\

\subsubsection{Spin-Dependent Scattering off Protons}
The sensitivity of target material depends on the presence of an unpaired nucleon in its atom. The target material used by PICO experiment $C_3 F_8$ features an unpaired proton, while xenon, the target material of LUX, features an unpaired neutron. Consequently the two experiments test individually the spin-dependent scattering off protons ($SD_p$) and neutrons ($SD_n$) respectively. \\

\noindent
In particular, the bubble chamber PICO-60 detector, which sets the strongest limits on $SD_p$, had an exposure of 1167 kg$\cdot$days of data taken between November 2016 and January 2017, excluding $SD_p$ of $3.4 \times 10^{-41} cm^2$ for a $30$~GeV DM mass \cite{Amole:2017dex}. One should notice that the limits presented by the PICO collaboration cover values of the DM mass only up to 1 TeV. Since we have considered, in our study, a broader mass range for the DM we have used an extrapolation. This operation is reliable since for heavy DM masses the scattering rate scales linearly with the number density of DM particles.\footnote{In reality, xenon detectors are also sensitive to SD scattering off protons. This is because the even number of proton in the xenon isotopes do not perfectly cancel each other's spin, giving rise to a small net spin. Such small net spin explains why xenon based detectors provide relatively much weaker limits on the SD dark matter scattering off protons. The current and projected limits from XENON1T experiment have been obtained but they are substantially weaker than PICO's and therefore were dropped out.}\\ 

\noindent
Dark Matter scattering on nuclei is not only tested in underground facilities. Searches of neutrino fluxes coming from annihilations of DM particles captured in the Sun can represent an even more powerful probe (see e.g.~\cite{Yaguna:2016bga}). We remind indeed that the flux of neutrinos observed from Sun is connected to the DM capture and annihilation rate at the Sun. The capture rate is mostly governed by the DM scattering off hydrogen, helium and oxygen, while the destruction rate is instead governed by the annihilation cross-section in the $v \rightarrow 0$ limit. In case that the equilibrium condition between the capture and destruction rate is met, it is possible to get rid of the dependence of the neutrino flux on the annihilation cross-section and cast limits only in terms of the SD cross-section. To this purpose is relevant to mention that the equilibrium condition can be satisfied for values of the annihilation cross-section much below the thermal value as long as the SD scattering cross section is sufficiently large. For this reason the model under study can be efficiently tested by neutrino telescopes while the typical values of the DM annihilation cross-section are not accessible to conventional indirect detection strategies. A residual dependence on the annihilation cross-section is nevertheless present since the neutrino flux actually depends on the type of annihilation final states since they induce distinct neutrino yields, and consequently different limits on the SD cross-section according to the dominant annihilation channel of the DM. This effect has been taken into account in our study.  For concreteness the DM capture rate at the Sun can be written as \cite{Ibarra:2014vya},

\begin{equation}
C_{DM} = 10^{20} s^{-1} \left( \frac{1~\text{TeV}}{m_\chi} \right)^2 \frac{2.77\, \sigma_{SD_p} + 4270\, \sigma_{SI_p} }{10^{-40}cm^{-2} }
\label{eq:Capture}
\end{equation}for DM masses above $1$~TeV.

\noindent
From \cref{eq:Capture} one can see that the non-observation of a neutrino signal from the Sun can yield limits on both the SD and SI scattering cross sections. The bounds on the SI are stronger due to larger overall factor in \cref{eq:Capture}. Currently IceCube imposes $SD_{p} < 10^{-40} cm^2$ and $SI_p < 10^{-43} cm^2$ for a $100$~GeV DM mass, annihilating into WW \cite{Aartsen:2016zhm}. Although, direct detection experiments such as XENON, LUX and PANDA-X provide already limits below $10^{-45} cm^2$ on SI scattering cross section \cite{Akerib:2015rjg,Fu:2016ega,Yang:2016odq,Tan:2016zwf,Aprile:2016swn}, while PICO sets $SD_{p} < 4\times10^{-41} cm^2$ \cite{Amole:2017dex}. For the this reason, IceCube searches for DM annihilations in the Sun are only truly relevant when it comes to $SD_p$ scattering. In the example above we used IceCube limits for DM annihilations into WW gauge bosons, but if we had adopted the bounds for annihilations into $\tau\tau$ which yield a harder neutrino spectrum, then IceCube limits can indeed be better then the one stemming from PICO, excluding $SD_p=2 \times 10^{-41} cm^2$.  Therefore, keep in mind that the limits coming from IceCube detector appearing in figures are based on the SD dark matter scattering off protons.

We now discuss the bounds based on SD scatterings off neutrons.

\subsubsection{Spin-Dependent Scattering off Neutrons}

As aforementioned the presence of an unpaired neutron in xenon isotopes makes xenon based detectors such as PANDA-X, LUX and XENON1T experiments the most sensitive to such nuclear recoil interactions.\\

\noindent
LUX collaboration has recently placed new limits on $SD_n$ using 129.5 kg-year exposure, excluding $SD_n =1.6 \times 10^{-41} cm^2$ for a $35$~GeV DM mass \cite{Akerib:2017kat}. An older limits has been casted by PANDA-X collaboration which is slightly weaker than LUX's \cite{Fu:2016ega}. This LUX limit is represented by a dashed green line. \\

\noindent
As for the projected sensitivity on $SD_n$ we adopted as baseline the XENON100 results. Since XENON1T collaboration expects to achieve a two orders of magnitude improvement on the SI cross section with 2 year$\times$ton exposure over the previous XENON100 result, we assumed the same rescalling for SD scattering cross section \citep{Aprile:2013doa}. In other words, since XENON1T projected limit with full exposure is expected to improve its predecessor by a factor of 100 concerning spin-independent scattering, we use this same 100 factor to project its sensitivity to spin-dependent DM-neutron scattering. \\

\noindent
Now we have described the experimental searches for spin-dependent DM-nucleon scattering we shall remark some key theoretical ingredients having in mind \cref{Eq:sigvff2,Eq:sigvZpZp,eq:SD}:\\

\noindent
{\bf (i)} Notice that the scattering cross section off nucleons scales with $Z^\prime$ mass to the forth power. Since we will fix the $g_\chi$ coupling to different values, the direct detection limits based on this scattering cross section will simply be straight lines in the Log-Log scale plots as shown in~\cref{fig:limits_g0,fig:limits_g1,fig:limits_g10}.\\

\noindent
{\bf (ii)} Since we have a sequential $Z^\prime$ gauge boson, $g_{ua}=g_{da}$, there will be no theoretical bias toward scattering off proton or neutrons.\\

\noindent
{\bf (iii)} PICO, LUX and XENON1T experimental sensitivities to our model rely only on the experimental parameters $\Delta_u^N,\Delta_d^N$ and $\Delta_s^N$.   \\

\noindent
{\bf (iv)} The annihilation into SM fermions might favor a particular final state depending on the DM mass, due to kinematic  effects. These threshold effects are rather visible near the top-quark and the $Z^\prime$ mass.  These channels have a significant impact on the annihilation cross section. These two effects explain the wavy behavior of the IceCube limits exhibited in \cref{fig:limits_g0,fig:limits_g1,fig:limits_g10}.


\begin{figure}[h]
\centering
\subfigure{
\includegraphics[height=0.11\textheight]{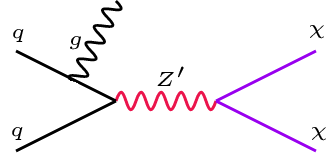} 
}
\subfigure{
\includegraphics[height=0.1\textheight]{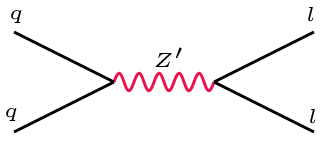} 
}
\caption{\label{fig:diagramCL} 
Feynman diagrams relevant for collider probes. The first diagram represents mono-jet searches for dark matter, where the $Z^\prime$ decays invisibly with a jet being radiated from the initial state. The second accounts or the resonance production of the $Z^\prime$ gauge boson. The latter is not particularly sensitive to dark matter, but it strongly restricts the $Z^\prime$ mass with great impact to this particular model. 
}
\end{figure}

\begin{figure}[!t]
\centering
\includegraphics[height=0.55\textheight]{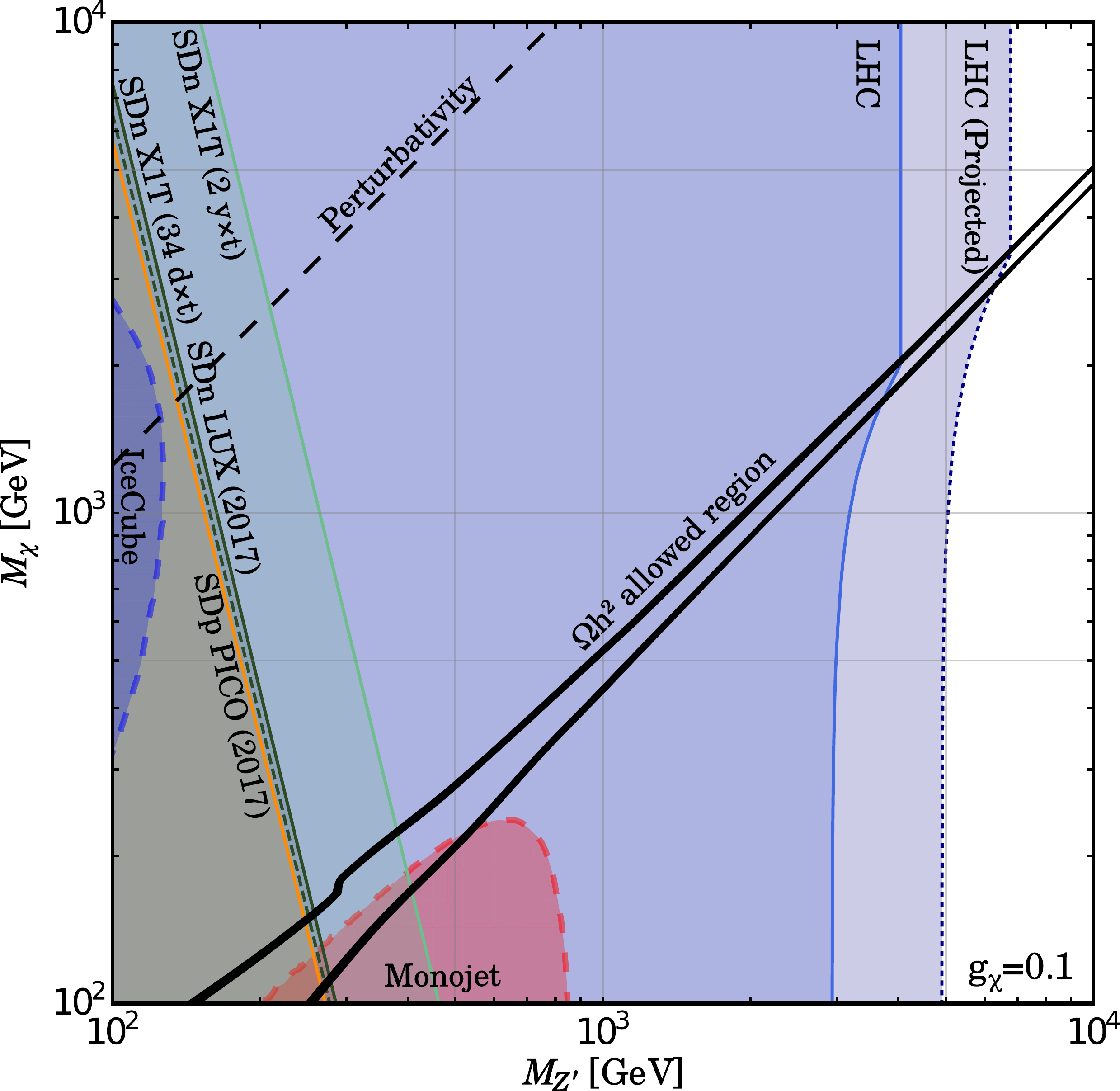}
\caption{\label{fig:limits_g0} 
Summary plot for $g_\chi=0.1$. The solid back curve outlines the region of parameter space with the correct relic density. From left to right: in dashed blue the parameter space excluded by IceCube; the orange solid line represents the current bound from PICO; the dashed green line the current bound from LUX on SD scattering off neutrons with 129.5 kg-year exposure; the solid green line the projected bound from XENON1T on SD scattering off neutrons with 34 d$\times$t of exposure; further right in light green, we show the projected sensitivity from XENON1T on SD scattering off neutrons with 2 y$\times$t exposure; the upper region inside the dashed black line delimits the non-perturbative regime; the dashed red curve exhibits the parameter space excluded by LHC based on mono-jet data; solid (dotted) blue vertical lines delimit the current (projected) LHC exclusion regions derived from dilepton data.}
\end{figure}

\begin{figure}[!t]
\centering
\includegraphics[height=0.55\textheight]{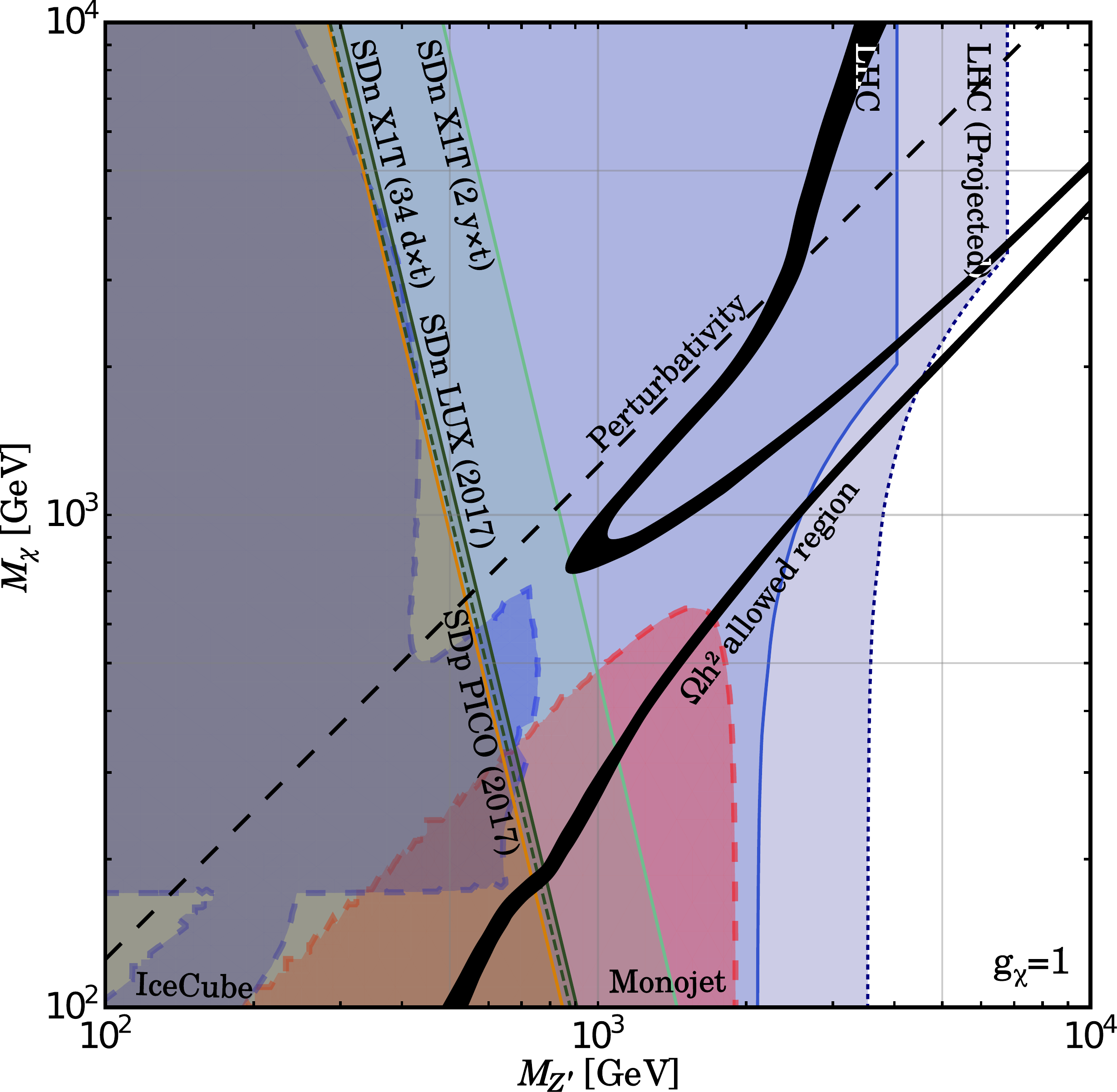}
\caption{\label{fig:limits_g1} 
Exclusion limits for $g_\chi=1$. The solid back curve outlines the region of parameter space with the correct relic density. From left to right: in dashed blue the parameter space excluded by IceCube; the orange solid line represents the current bound from PICO; the dashed green line the current bound from LUX on SD scattering off neutrons with 129.5 kg-year exposure; the solid green line the projected bound from XENON1T on SD scattering off neutrons with 34 d$\times$t of exposure; further right in light green, we show the projected sensitivity from XENON1T on SD scattering off neutrons with 2 y$\times$t exposure; the upper region inside the dashed black line delimits the non-perturbative regime; the dashed red curve exhibits the parameter space excluded by LHC based on mono-jet data; solid (dotted) blue vertical lines delimit the current (projected) LHC exclusion regions derived from dilepton data.
}
\end{figure}

\begin{figure}[!t]
\centering
\includegraphics[height=0.55\textheight]{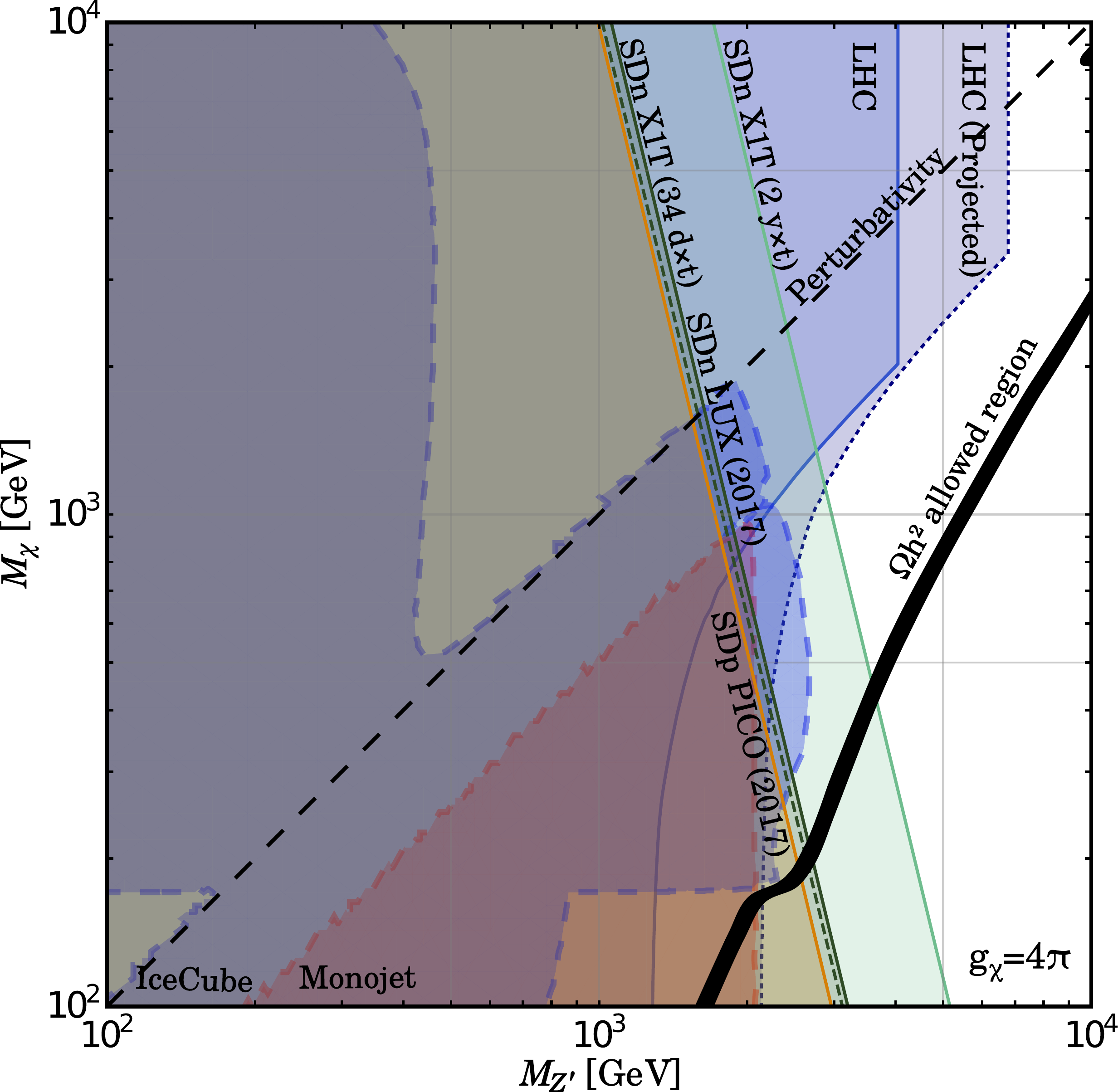}
\caption{\label{fig:limits_g10} 
Exclusion limits for $g_\chi=4\pi$. The solid back curve outlines the region of parameter space with the correct relic density. From left to right: in dashed blue the parameter space excluded by IceCube; the orange solid line represents the current bound from PICO; the dashed green line the current bound from LUX on SD scattering off neutrons with 129.5 kg-year exposure; the solid green line the projected bound from XENON1T on SD scattering off neutrons with 34 d$\times$t of exposure; further right in light green, we show the projected sensitivity from XENON1T on SD scattering off neutrons with 2 y$\times$t exposure; the upper region inside the dashed black line delimits the non-perturbative regime; the dashed red curve exhibits the parameter space excluded by LHC based on mono-jet data; solid (dotted) blue vertical lines delimit the current (projected) LHC exclusion regions derived from dilepton data.
}
\end{figure}

\subsection{Colliders}

Since we are discussing a model in the context of vector mediators, it is well known that the most efficient way to probe this simplified dark matter model is through the mono-jet and dilepton data sets \cite{Alves:2013tqa}, see \cref{fig:diagramCL}. The latter proving the most restrictive limits.\\

\noindent
The limits we included were derived from the LHC searches for the Sequential SM $Z^\prime$ decaying into charged leptons \citep{Aaboud:2017buh} with an integrated luminosity of $36.1fb^{-1}$ and $13$~TeV center-of-mass-energy. They rely on the fact that a heavy spin-1 additional boson $Z^\prime$ would decay producing a narrow resonance in the dilepton channel that has a low and well-understood background. Subsequently an upper limit on the cross-section times the branching ratio can be extracted, and consequently a lower bound on the $Z^\prime$ mass can be derived. Two limits are often quoted. One based on the dielectron data ($4.3$~TeV) and other in the dimuon data ($4$~TeV). The dielectron channel yields a slightly stronger limit than the dimuon one due to the larger acceptance/efficiency \cite{Aaboud:2017buh}. The sequential $Z^\prime$ boson features a bit large decay width and therefore the bound on the $Z^\prime$ mass is subject to sizable uncertainties. For instance, in the combined channel (dielectron $+$ dimuon) the lower mass bound on the $Z^\prime$ ranges from $4.3$~TeV to $4.8$~TeV (see Fig.4 of \citep{Aaboud:2017buh}). Moreover, there are also mild systematic uncertainties associated to dilepton resonant searches \cite{Accomando:2016ehi}. In light of that, we decided to take a conservative approach and adopted the LHC limit obtained using the dimuon channel, which imposes $M_{Z^\prime} > 4$~TeV. This limit is represented by a solid vertical line in \cref{fig:limits_g0,fig:limits_g1,fig:limits_g10}. Despite applying the limit on the $Z^\prime$ mass based on dimuon data only, we refer to this collider bound as dilepton hereafter.\\

\noindent
We also include the projected bound under the null result hypothesis for center-of-mass-energy of $14$~TeV and integrated-luminosity of 1000 fb$^{-1}$~\footnote{ \url{http://collider-reach.web.cern.ch/?rts1=13&lumi1=3.2&rts2=13&lumi2=13.3&pdf=MSTW2008nnlo68cl.LHgrid}} as a dotted blue line in~\cref{fig:limits_g0,fig:limits_g1,fig:limits_g10} that  would rule out the region $M_{Z^\prime}\approx 6.7$ TeV. 

\noindent
As one clearly sees the exclusion limits in \cref{fig:limits_g0,fig:limits_g1,fig:limits_g10}, from searches of dilepton resonances, become weaker in the lower portion of the plots, i.e. when the DM is lighter than the $Z'$. Indeed the analysis from the ATLAS experiment incorporate only decay channels into SM states for the $Z'$. This does not occur in our case when $2 m_\chi < M_{Z'}$. The total decay width of the $Z'$ gets indeed modified as \cite{Arcadi:2013qia},
\begin{equation} \label{eq:Gamma}
\begin{aligned}
\Gamma_{Z^\prime} = &\sum_{f\in\text{SM}}\theta(M_{Z^\prime}-2m_f)\dfrac{n_c M_{Z^\prime}}{12\pi}\sqrt{1-\dfrac{4m_f^2}{M_{Z^\prime}^2}}\left[ g_{fv}^2\left( 1+\dfrac{2m_f^2}{M_{Z^\prime}^2}\right) + g_{fa}^2\left( 1-\dfrac{4m_f^2}{M_{Z^\prime}^2}\right)\right]\\
& \theta(M_{Z^\prime}-2m_\chi)\dfrac{M_{Z^\prime}}{12\pi}\sqrt{1-\dfrac{4m_\chi^2}{M_{Z^\prime}^2}}g_{\chi}^2\left( 1-\dfrac{4m_\chi^2}{M_{Z^\prime}^2}\right),
\end{aligned}
\end{equation}
where $g_{fv}$ and $g_{fa}$ were given in \cref{eq:g} and $\theta$ is the unit step function.  Consequently, the branching ratio to dilepton becomes in this regime,
\begin{equation}
\begin{aligned}
\dfrac{\Gamma(Z^\prime\to \ell\ell)}{\Gamma(Z^\prime\to ff)}& \Rightarrow \dfrac{\Gamma(Z^\prime\to \ell\ell)}{\Gamma(Z^\prime\to ff) + \Gamma(Z^\prime\to \chi\chi)}\\
& = \dfrac{\Gamma(Z^\prime\to \ell\ell)}{\Gamma(Z^\prime\to ff)}\left(1-\text{Br}(Z^\prime\to \chi\chi)\right)=\text{Br}(Z^\prime_{\rm{SSM}}\to \ell\ell)\left[1-\text{Br}(Z^\prime\to \chi\chi)\right]
\end{aligned}
\end{equation}where $f$ is a SM fermion.\\

\noindent
Therefore, the exclusion limit on the $Z^\prime$ mass which depends linearly on $\text{Br}(Z^\prime_{\rm{SSM}}\to \ell\ell)$ will be weakened by $\left[1-\text{Br}(Z^\prime\to \chi\chi)\right]$. Obviously, this effect takes place only when the decay of the $Z'$ into DM pairs is kinematically accessible, as aforementioned. As can be easily argued the effect of opening the invisible decay channel is more prominent at the highest values of the coupling $g_\chi$ since they correspond to higher values of the invisible branching fraction of the $Z'$.\\

\noindent
The monojet bound features a complementary behavior with respect to the dilepton one. It is indeed based on searches of monojet events plus missing energy whose production rate is maximal when the $Z'$ decay on-shell mostly on DM pairs. For this reason the strongest bound is obtained for $g_\chi=4 \pi$ and $m_\chi < M_{Z'}/2$. On the contrary, the size of the excluded region is increasingly reduced as $g_\chi$ decreases and substantially no bound is present for $m_\chi > M_{Z'}/2$.

\subsection{Perturbativity}
\label{ssec:perturbativity}

\noindent
As already discussed the annihilation cross-section associated to $\chi \chi \rightarrow Z'Z'$ process show a rather peculiar behavior, scaling, for $m_\chi \gg M_{Z'}$, as $m_{\chi}^2/M_{Z'}^4 v^2$ and then indefinitely increasing with the value of the DM mass. As discussed in~\cite{Shu:2007wg,Kahlhoefer:2015bea} this is due to the longitudinal degrees of freedom of the $Z'$ which induce a contribution in the annihilation amplitude proportional to $\sqrt{s} m_\chi/M_{Z'}^2$ (the annihilation into longitudinal degrees of freedom would actually induce a $s/M_{Z'}^2$ scaling. The dependence on s is weakened because of cancellation between t- and u- channel diagrams~\cite{Kahlhoefer:2015bea}). The behavior of the annihilation cross-section is, at this point, easily understood once remembering that for the relic density only the non-relativistic limit, $s \sim 4 m_\chi^2$, is relevant. The fact that the contribution associated to the annihilation into longitudinal $Z'$ pair appears in the p-wave term of velocity expansion can be inferred through $CP$ and angular momentum conservation arguments~\cite{Drees:1992am}. As widely known amplitudes increasing with the center of mass energy are pathological and violate perturbative unitarity at relatively low energy. The presence of a unitarity violation cross-section is caused by the fact that we are considering a not UV complete framework. In $Z'$ models based on the spontaneous breaking of extra gauge symmetries the annihilation rate into $Z'Z'$ is cured once the diagram with s-channel exchange of the scalar field responsible for the spontaneous breaking of the new theory is accounted for (see e.g.~\cite{Duerr:2016tmh}. Discussions on similar lines can be found also in~\cite{Bell:2016uhg,Cui:2017juz}).\\

\noindent
Since we do not rely on a UV complete model, we will apply a limit from not violation of perturbative unitarity in the process $\chi \chi \rightarrow Z'Z'$, in the same fashion as~\cite{Kahlhoefer:2015bea}:
\begin{equation}
\sqrt{s}>\frac{\pi M_{Z^\prime}}{g_\chi^2 M_\chi},
\end{equation}
which can be expressed, in the non-relativistic limit relevant for the DM relic density as:
\begin{equation}\label{eq:pert}
{M_\chi}>\sqrt{\frac{\pi M_{Z^\prime}}{2g_\chi^2}}.
\end{equation}

We emphasize that \cref{eq:pert} should be interpreted as a limit for the validity of the computations presented in this work. Beyond the region of parameter space delimited by \cref{eq:pert} one should explicitly take into account additional degrees of freedom needed to unitarize the theory. \\

\noindent
This condition excludes the yellow region in~\cref{fig:limits_g0,fig:limits_g1,fig:limits_g10}. Now that we have described all observables of the simplified Majorana dark matter model, we will gather all ingredients and comment on our findings.

\section{Results}
\label{sec:res}

\noindent
Our main results are summarized in~\cref{fig:limits_g0,fig:limits_g1,fig:limits_g10} in the bidimensional plane $M_{Z'}, m_\chi$. In the plot the parameter space accounting for the correct DM relic density are compared with the limits from the most relevant dark matter observables taking into account current and future experimental sensitivities to outline the region of parameter space where one can have a viable Majorana fermion as dark matter. The individual origin of the different bounds reported in the plot has been discussed in the previous sections. Here we will discuss more extensively the effect of their combination and the impact on the parameter space.\\

\noindent
We start by commenting on the relic density. The correct DM relic density is represented, in~\cref{fig:limits_g0,fig:limits_g1,fig:limits_g10}, by black isocontours. For the lowest assignation of $g_\chi$, namely 0.1, the correct relic density is achieved only through resonantly enhanced, for $m_\chi \sim m_{Z'}/2$, annihilation into SM fermions. At $g_\chi=1$ the correct relic density is also easily achieved, for $m_\chi > M_{Z'}$ through the $\chi \chi \rightarrow Z'Z'$ annihilation process, and also a bit far from the resonance, when annihilations into $\bar t t$ are maximally efficient. A large part of the viable parameter space for $m_\chi > M_{Z'}$ is, however, excluded by the unitarity constraint. For $g_\chi=4 \pi$ finally, the annihilation into $Z'Z'$ is too efficient, always leading to under abundant DM, and the correct relic density is achieved through annihilations into SM fermions far from the $m_\chi \sim M_{Z'}/2$ pole region. \\

\noindent
Concerning the limit from LHC, for fixed values of the couplings, the limits from searches of dilepton resonances, as long as $M_{Z'}< 2 m_\chi$, they are basically turned into an exclusion in the mass of the $Z'$ independent on the value of $g_\chi$. On the contrary, when $M_{Z'}> 2 m_\chi$ the excluded value of $M_{Z'}$ suffers the rescaling effect, described before, attributed to the invisible branching fraction of the $Z'$ and actually depends on $m_\chi,g_\chi$. For $g_\chi=4 \pi$ the exclusion bound can be reduced to 1 TeV~\footnote{Notice that in this context one should consider a complementary bound from LEP~\cite{Langacker:2008yv} coming from eventual modifications of the dielectron production cross-section. This kind of search tests the off-shell production of the $Z'$ and then the limits depend only on its mass and coupling with the electrons, irrespective of the presence of other couplings. For the Sequential $Z'$ the limit is of approximately $1.8\,\mbox{TeV}$. For simplicity we have not reported the corresponding line in the plots. See however~\cite{Arcadi:2017kky}.} while for the lowest assignation $g_\chi$ the effect of the invisible branching ratio is almost marginal.\\

Generally speaking the bound from dilepton searches is the strongest for the kind of scenario under consideration (see also~\cite{Arcadi:2017atc} for a dedicated study). The only exception is represented by the extreme assignation $g_\chi$ for which, in the region $m_\chi < M_{Z'}/2$ DM Direct Detection poses the most competitive constraints. Although in principle complementary to searches searches of dilepton resonances, searches for monojet events are not yet competitive with respect to other observational constraints. The reason mostly lies in the larger SM backgrounds which plague monojet searches with respect to the ones of lepton final states. 

\noindent
The sensitivity of this kind of experimental searches will receive a substantial improvement from XENON1T. Constraints by IceCube demonstrate a potentially good complementarity. They are however strongly dependent by the kind of annihilation final state of the DM. In the model considered they are mostly effective in the intermediate mass range both the $Z'$ and the DM, where the latter dominantly annihilates into $\bar t t$.

\noindent
A final remark that is important to make. As emphasized, our findings are valid in the context of the sequential $Z^\prime$ model. However, since the direct detection limits scale with $g^2_{\chi} g_{fa}^2$ as well as the annihilation rate into SM fermions, and the dilepton bounds roughly scale up with $g_{fv}^4$, one can recast one findings to many Majorana fermion dark matter models. Furthermore, we emphasize that our conclusions rely on thermal production of dark matter and standard cosmology. Departure from these two assumptions would consequently change the relic density curves and the quantitative assessments based on the latter.

\section{Conclusions}
\label{sec:conclusions}

We have investigated the Majorana dark matter model in the context of the $Z^\prime$ portal. The dark matter phenomenology is dictated by gauge interactions which are fixed, since we adopted the Sequential $Z^\prime$ framework, rendering our simplified model predictive. Direct dark matter detection experiments based on bubble chamber and liquid xenon, as well as neutrino telescopes observing neutrinos from the Sun provide complementary tests to this model. LHC searches for {\it dilepton} and {\it mono-jet + missing energy} events provide orthogonal bounds to the parameter space. The former being the stronger one. We computed the relic density curves and outlined the region of parameter space where one can successfully have a Majorana dark matter particle in agreement with data. \\

\noindent
We varied the dark matter coupling to the $Z^\prime$ to asses the impact on the constraints and highlight the importance of complementary probes for dark sectors.\\

\noindent
In summary, the Majorana dark matter fermion model via the $Z^\prime$ portal offers a gripping dark matter phenomenology with exciting implications to neutrino detectors, underground direct detection experiments as well as colliders. Therefore, it should be treated as a benchmark model in dark matter research endeavors.

\section*{Acknowledgments}

The authors thank Carlos Yaguna, Stefan Vogl, Werner Rodejohann, Clarissa Siqueira, Maira Dutra, Alex Dias for discussions. FSQ thanks Yann Mambrini and Abdelhak Djouadi from Orsay-LPT for the hospitality during the final stages of this project. M. C. was supported by the IMPRS-PTFS.
The research of A.M. is supported by the ERC Advanced Grant No.
267985 (DaMeSyFla) and by the INFN. A.M. gratefully acknowledges support by the research grant Theoretical Astroparticle Physics No. 2012CPPPYP7 under the program PRIN 2012 funded by the MIUR.

\bibliographystyle{JHEPfixed}
\bibliography{MajDM}

\end{document}